\documentclass{article}
\usepackage{latexsym}
\usepackage{amssymb}
\usepackage{amsmath}
\usepackage[numbers,sort&compress]{natbib}
\usepackage{color}
\usepackage{amsfonts,amssymb}
\usepackage{mathrsfs}
\usepackage{dsfont}
\usepackage{graphicx}

\newtheorem{remark}{Remark}[section]
\newtheorem{prop}{Proposition}[section]
\def\cdot{{\scriptstyle\,\bullet\,}}

\begin{document}
\title{\bf On an integrable system related to the relativistic Toda lattice -B\"acklund
transformation and integrable discretization}
\author{Luc Vinet$^{\dag}$, Guo-Fu Yu$^{\ddag}$
\footnote{Corresponding author: gfyu@sjtu.edu.cn } and Ying-Nan Zhang$^{\S,\sharp}$\\
\\ $^{\dag}$
Centre de Recherches Math\'{e}matiques, Universit\'{e} de
Montr\'{e}al,
\\ C.P.6128, Centre-ville Station, Montr\'{e}al, Qu\'{e}bec, H3C 3J7, Canada\\
$^{\ddag}$Department of Mathematics, Shanghai Jiao Tong University, \\
Shanghai 200240, P.R.\ China
\\ $^{\S}$LSEC, Institute of Computational Mathematics and Scientific
Engineering Computing,\\
Academy of Mathematics and System Sciences,\\
Chinese Academy of Sciences, P.O.\ Box 2719, Beijing 100080, P.R.\
China \\
$^{\sharp}$Graduate School of the Chinese Academy of Sciences,
Beijing, P.R.\ China \\ }
\date{}
\maketitle

\begin{abstract}
We study an integrable system related to the relativistic Toda
lattice. The bilinear representation of this lattice is given and
the B\"ackulund transformation obtained.  A fully discrete version
is also introduced with its bilinear B\"acklund transformation and
Lax pair. One-soliton solution of the discrete system is presented
by use of B\"acklund transformation.
\end{abstract}

\section{Introduction}
In terms of the canonical variables $\{q_n,\Theta_n\}_{n\in Z}$, the
Hamiltonian for the infinite relativistic Toda  lattice (RTL) is
given by \cite{Bruschi,Ragnisco1}
\begin{eqnarray}
H(q,\Theta)=\sum_{n\in
Z}\{\exp(\Theta_n)[1+\exp(q_{n-1}-q_n)]^{1/2}\times
[1+\exp(q_n-q_{n+1})]^{1/2}-2 \}.
\end{eqnarray}
The equations of motion are hence
\begin{align}
& \frac{d}{dt}q_n=\frac{\partial H_n}{\partial \Theta_n}=b_n,\\
& \frac{d}{dt}\Theta_n=-\frac{\partial H_n}{\partial
q_n}=\frac{1}{2}a_{n-1}(b_{n}+b_{n-1})-\frac{1}{2}a_n(b_n+b_{n+1}),
\end{align}
with
\begin{eqnarray}
&& b_n=\exp(\Theta_n)[1+\exp(q_{n-1}-q_n)]^{1/2}\times
[1+\exp(q_n-q_{n+1})],\\
&& a_n=\frac{\exp(q_n-q_{n+1})}{1+\exp(q_n-q_{n+1})}.
\end{eqnarray}
The RT equation
\begin{align}
\ddot{q}_n=&(1+\frac{1}{c}\dot{q}_{n-1})(1+\frac{1}{c}\dot{q}_n)
\frac{\exp(q_{n-1}-q_n)}{1+(1/c^2)\exp(q_{n-1}-q_n)}\nonumber\\
&-(1+\frac{1}{c}\dot{q}_{n})(1+\frac{1}{c}\dot{q}_{n+1})
\frac{\exp(q_{n}-q_{n+1})}{1+(1/c^2)\exp(q_{n}-q_{n+1})},\label{ruij}
\end{align}
where $q_n$ is the coordinates of $n-$th lattice point, and means
the differentiation with respect to time $t$ and $c$ is the light
speed, was introduced and studied by Ruijsenaars \cite{ruij90}. The
evolution equations of $\{a_n, b_n\}$ assume the form
\begin{eqnarray}
&& \frac{d}{dt}b_n=b_n(b_{n-1}a_{n-1}-a_nb_{n+1}),\label{a1}\\
&& \frac{d}{dt}a_n=a_n(1-a_{n})(b_n-b_{n+1}).\label{a2}
\end{eqnarray}
By a further transformation
\begin{eqnarray}
&& u_n=b_n(1-a_n),\quad v_n=a_nb_n,
\end{eqnarray}
the equations \eqref{a1}-\eqref{a2} are rewritten as
\begin{eqnarray}
&& \frac{d}{dt}u_n=u_n(v_{n-1}-v_n),\label{rt1}\\
&& \frac{d}{dt}v_n=v_n(v_{n-1}-v_{n+1}+u_n-u_{n+1}).\label{rt2}
\end{eqnarray}
We refer to \eqref{rt1}-\eqref{rt2} as the RTL.
\cite{Ragnisco1,Common1,Suris1997}.

The integrable lattice equations are related to the following
discrete spectrum problem and time evolution equation
\begin{eqnarray}
&& E\varphi_n=U_n(u,\lambda)\varphi_n, \label{d1}\\
&& \varphi_{n,t}=V_n(u,\lambda)\varphi_n,\label{d2}
\end{eqnarray}
where $u$ is the potential function, $\lambda$ the spectral
parameter and $E$ the shift operator defined by $Ef_n=f_{n+1}$.

The integrability condition between \eqref{d1} and \eqref{d2} leads
to the integrable lattice system
\begin{eqnarray}
U_{n,t}+U_nV_n-V_{n+1}U_n=0,\label{cd}
\end{eqnarray}
when we take $u=(u_n,v_n)^{T}$, and
\begin{eqnarray}
&& U_n= \left(\begin{matrix}
\lambda^2+u_n & \lambda \\
\lambda v_n & 0
\end{matrix}\right)\label{uex}\\
&& V_n=\left(\begin{matrix}
\lambda^2/2-v_{n-1} & \lambda \\
\lambda v_{n-1} & -\lambda^2/2-u_n-v_n
\end{matrix}\right).\label{vex}
\end{eqnarray}
The compatibility condition \eqref{cd} entails the RTL equation
\eqref{rt1}-\eqref{rt2}. So \eqref{d1} and \eqref{d2} with
\eqref{uex}-\eqref{vex} constitute a Lax pair for the RTL
\eqref{rt1}-\eqref{rt2}. Other Lax presentation for
\eqref{rt1}-\eqref{rt2} can be found in \cite{Suris1997,
ma2004,fan2008,Wenxy}. The bilinear form and Casorati determinant
solution for the RTL \eqref{ruij} were given in \cite{ohta1993}. It
is also worth pointing out that the RTL \eqref{rt1}-\eqref{rt2} can
be related to the Laurent bi-orthogonal polynomials \cite{zh,Vinet}.
The discretization of the RTL \eqref{ruij} was first performed in
\cite{suris1996} using the Hamiltonian method. The discrete-time
relativistic Toda lattice (dRTL) equation was first proposed by
Suris
\begin{align}
\frac{\delta \exp(q_n^{t+1}-q_n^t)-1}{\delta
\exp(q_n^{t}-q_n^{t-1})-1}=\frac{1+g^2\exp(q_{n-1}^{t}-q_n^t)}
{1+g^2\exp(q_{n}^{t}-q_{n+1}^t)}\frac{1+(g^2/\delta)\exp(q_n^t-q_{n+1}^{t+1})}
{1+(g^2/\delta)\exp(q_{n-1}^{t-1}-q_{n}^{t})}.\label{drt}
\end{align}
The Lagrangian form of dRTL was presented in \cite{zhu99}. The
Casorati determinant solution for the discrete RT lattice
\eqref{drt} was given in \cite{Maruno1998} and the elliptic
solutions in \cite{Tsu_Alex}.

Although the discrete version of the RTL \eqref{ruij} is already
known, as far as we know, the discrete analogue of the RTL related
system \eqref{rt1}-\eqref{rt2} has not been given. The purpose of
this paper is to propose a direct discrete version of the RTL
\eqref{rt1}-\eqref{rt2} using Hirota bilinear method. It is
nontrivial and of considerable interest to find integrable
discretizations for integrable equations. Attention is being paid to
the problem of integrable discretizations of integrable systems.
(See e.g., \cite{LVW,CN,LR,HNS,suris} and references therein).
Various approaches to the problem of integrable discretization are
currently available. One of them is Hirota's bilinear method
\cite{Hd1,Hd2,Hd3,Hd4,Hd5,H1}, which is  based on gauge invariance
and soliton solutions. Here, we focus on a discretization process
such that the resulting discrete bilinear equations have B\"acklund
transformations (BTs). As a bonus, we can usually derive Lax pairs
for the resulting discrete equations. This method has been
successfully applied to the discretization of $(2 + 1)$-dimensional
sinh-Gordon equation \cite{Huyu1} and of the two dimensional Leznov
lattice equation \cite{Huyu2}. Based on bilinear forms and
determinant structure of solutions, Hirota's discretization method
has also been developed to construct discrete versions of the
Camassa-Holm equation \cite {ch} and the short pulse equation \cite
{sp}.

The content of the paper is organized as follows. In section $2$, we
give a bilinear form of the RTL. In section $3$, a BT and Lax pair
are exhibited. In section $4$ an integrable discretization of the
RTL is presented and the integrability is made manifest by the
corresponding BT. Section $5$ is devoted to conclusion and
discussions.

\section{Bilinear form for the RTL}
\setcounter{equation}{0} Performing the change of the dependent
variable
\begin{eqnarray}
v_n=\frac{d}{dt}\ln \frac{F_n}{G_n}+1,\qquad u_n=\frac{d}{dt}\ln
\frac{G_{n-1}}{F_n}+1,
\end{eqnarray}
the system (\ref{rt1})-(\ref{rt2}) is transformed into the following
bilinear form
\begin{eqnarray}
&& D_t G_{n-1}\cdot F_n +F_nG_{n-1}=c_1 F_{n-1}G_n,\label{brt1}\\
&& D_t F_n\cdot G_n+F_nG_n=c_2F_{n-1}G_{n+1},\label{brt2}
\end{eqnarray}
which can also be written as follows in terms of the Hirota bilinear
differential:
\begin{eqnarray}
&& (D_te^{\frac{1}{2}D_n}-e^{\frac{1}{2}D_n}+c_1e^{-\frac{1}{2}D_n}) F \cdot G=0,\\
&& (D_t+1-c_2e^{-D_n})F\cdot G=0.
\end{eqnarray}
Here $c_1, c_2$ are arbitrary constants. In the following we shall
take $c_1=c_2=1$ for simplicity. The Hirota bilinear differential
operator $D_t^k$ and the bilinear difference operator $\exp(\delta
D_n)$ are respectively defined \cite{RH1} by
\begin{equation*}
D_t^k a\cdot b\equiv \left (\frac \partial {\partial t}-\frac
\partial {\partial t'}\right )^ka(t)b(t')|_{t'=t},
\end{equation*}
\begin{equation*}
\exp(\delta D_n)a(n)\cdot b(n)\equiv a(n+\delta)b(n-\delta).
\end{equation*}
It is well known that nonlinear integrable equations share many
common features, among which, the BT and their associated nonlinear
superposition formulae \cite{Miura,RS1,RS2,Hu}.

\section{Bilinear B\"acklund transformation for the RTL}
\setcounter{equation}{0} For the sake of convenience, we introduce
an additional discrete variable $m$ and set
\begin{eqnarray}
F=f_{m+\frac{1}{2}}, \quad G=f_{m-\frac{1}{2}} \label{dvt}
\end{eqnarray}
We shall denote $F(t,n)$ by $F_n$ and $f(t,n,m)$ by $f_{n,m}$ for
simplicity. With the help of \eqref{dvt}, the eqs. $(\ref{brt1}),
(\ref{brt2})$ reduce to
\begin{eqnarray}
&& (D_t e^{\frac{D_n+D_m}{2}}-e^{\frac{D_n+D_m}{2}}+e^{\frac{D_m-D_n}{2}})f\cdot f=0,\label{brt3}\\
&& (D_t e^{\frac{D_m}{2}}+ e^{\frac{D_m}{2}}-
e^{\frac{D_m}{2}-D_n})f\cdot f=0.\label{brt4}
\end{eqnarray}

\begin{prop}
The bilinear equations (\ref{brt3}), (\ref{brt4}) have the
B\"acklund transformation
\begin{eqnarray}
&& (D_t-\lambda e^{-D_n}+\mu )f\cdot g=0,\label{bt1}\\
&&(e^{\frac{D_n-D_m}{2}}-e^{\frac{D_m-D_n}{2}}-\lambda
e^{-\frac{D_n+D_m}{2}})f \cdot g =0,\label{bt2}\\
&& (\lambda e^{-D_n-\frac{D_m}{2}}+e^{-\frac{D_m}{2}}+\gamma
e^{\frac{D_m}{2}})f\cdot g=0,\label{bt3}
\end{eqnarray}
where $\lambda, \mu$ and $\gamma$ are arbitrary constants.
\end{prop}
{\bf Proof.} Let $f$ be a solution of equations
\eqref{brt3}-\eqref{brt4}. If it can be shown that $g$ given by
\eqref{bt1}-\eqref{bt3} satisfies
\begin{eqnarray*}
&& P_1\equiv  (D_t e^{\frac{D_n+D_m}{2}}-e^{\frac{D_n+D_m}{2}}
+e^{\frac{D_m-D_n}{2}})g\cdot g=0,\\
&& P_2\equiv (D_t e^{\frac{D_m}{2}}+ e^{\frac{D_m}{2}}-
e^{\frac{D_m}{2}-D_n})g \cdot g=0,
\end{eqnarray*} then the equations \eqref{bt1}-\eqref{bt3}
define a BT for \eqref{brt3}-\eqref{brt4}.

In fact, using \eqref{bt1}-\eqref{bt3} and the bilinear identities,
it can be seen that

\begin{align*}
&-[e^{\frac{D_m}{2}}f\cdot f]P_2 \\
&\quad\equiv [(D_t e^{\frac{D_m}{2}}+ e^{\frac{D_m}{2}}-
e^{\frac{D_m}{2}-D_n})f \cdot f][e^{\frac{D_m}{2}}g\cdot g]- [(D_t
e^{\frac{D_m}{2}}+ e^{\frac{D_m}{2}}-
e^{\frac{D_m}{2}-D_n})g \cdot g][e^{\frac{D_m}{2}}f\cdot f]\\
&\quad =2\sinh(\frac{D_m}{2})[D_t f\cdot g]\cdot
(fg)-2\sinh(\frac{D_n}{2})(e^{\frac{D_n-D_m}{2}}f\cdot g)\cdot
(e^{\frac{D_m-D_n}{2}}f\cdot g)\\
& \quad =2\sinh(\frac{D_m}{2})[D_t f\cdot g]\cdot
(fg)+2\lambda \sinh(\frac{D_m}{2})(fg)\cdot (e^{-D_n}f\cdot g) \\
&\quad=2\sinh(\frac{D_m}{2})[(D_t-\lambda e^{-D_n})f\cdot g]\cdot (fg)\\
&\quad =0\\
&\\
&-[e^{\frac{D_m+D_n}{2}}f\cdot f]P_1 \\
&\quad\equiv [(D_t e^{\frac{D_n+D_m}{2}}-e^{\frac{D_n+D_m}{2}}
+e^{\frac{D_m-D_n}{2}})f\cdot f][e^{\frac{D_m+D_n}{2}}g\cdot g]\\
&\qquad\qquad\qquad\qquad - [(D_t
e^{\frac{D_n+D_m}{2}}-e^{\frac{D_n+D_m}{2}}
+e^{\frac{D_m-D_n}{2}})g\cdot g][e^{\frac{D_m+D_n}{2}}f\cdot f]\\
&\quad =2\sinh(\frac{D_m+D_n}{2})(D_tf\cdot g)\cdot
(fg)-2\sinh(\frac{D_n}{2})(e^{\frac{D_m}{2}}f\cdot g)\cdot
(e^{-\frac{D_m}{2}}f\cdot g)\\
&\quad =2\sinh(\frac{D_m+D_n}{2})(\lambda e^{-D_n}f\cdot g)\cdot
(fg)-2\sinh(\frac{D_n}{2})(e^{\frac{D_m}{2}}f\cdot g)\cdot
(e^{-\frac{D_m}{2}}f\cdot g)\\
&\quad =2\sinh(\frac{D_n}{2})(e^{\frac{D_m}{2}}f\cdot g)\cdot
[(-\lambda e^{-D_n-\frac{D_m}{2}}-e^{-\frac{D_m}{2}})f\cdot g]\\
&\quad =0.
\end{align*}

This completes the proof of proposition 2.1.

From Proposition 2.1 it can be deduced that the bilinear RTL
\eqref{brt3}-\eqref{brt4} has the following B\"acklund
transformation:
\begin{eqnarray}
&& D_t F \cdot
F'-\lambda F_{n-1}F_{n+1}'+\mu FF'=0,\label{bt21} \\
&& D_t G \cdot
G'-\lambda G_{n-1}G_{n+1}'+\mu GG'=0,\label{bt22}\\
&& G_{n+1}F'-FG_{n+1}'-\lambda GF_{n+1}'=0,\label{bt23}\\
&& \lambda G_{n-1}F_{n+1}'+GF'+\gamma FG'=0,\label{bt24}
\end{eqnarray}
where $F'$ and $G'$ are defined by
\begin{eqnarray}
F'=g_{m+\frac{1}{2}},\qquad G'=g_{m-\frac{1}{2}}.
\end{eqnarray}
Starting from the bilinear BT (\ref{bt21})-(\ref{bt24}), we can
derive a Lax pair for the system (\ref{rt1})-(\ref{rt2}). First, set
\begin{eqnarray*}
\Phi_{n}=\frac{F_n}{F'_n},\quad \Psi_n=\frac{G_n}{G'_n},\quad
P_{n}=\ln\frac{G'_{n-1}}{F'_{n}},\quad
Q_{n}=\ln\frac{F'_{n}}{G'_{n}}
 \end{eqnarray*}
in (\ref{bt21})-(\ref{bt24}). Eqs (\ref{bt21})-(\ref{bt24}) are then
transformed into
\begin{eqnarray}
&& \Psi_{n,t}-\lambda
e^{P_n+Q_n-P_{n+1}-Q_{n+1}}\Psi_{n-1}+\mu\Psi_n=0,\label{btU1}\\
&& \Phi_{n,t}-\lambda e^{P_n+Q_{n-1}-P_{n+1}-Q_n}\Phi_{n-1}+\mu
\Phi_n=0,\\
&& \Psi_{n+1}-\Phi_n-\lambda e^{Q_{n+1}-Q_n}\Psi_n=0,\\
&& \Psi_n+\gamma\Phi_n+\lambda
e^{P_n-P_{n+1}}\Psi_{n-1}=0.\label{btU2}
\end{eqnarray}
From the compatibility of above equations we now obtain
\begin{eqnarray}
&& \frac{d}{dt}P_{n}-e^{Q_{n-1}-Q_n}=\frac{d}{dt}P_{n+1}-e^{Q_{n}-Q_{n+1}},\label{2.7}\\
&&\frac{d}{dt}Q_n-e^{P_n-P_{n+1}+Q_{n-1}-Q_{n+1}}
=\frac{d}{dt}Q_{n+1}-e^{P_{n+1}-P_{n+2}+Q_{n}-Q_{n+2}},\label{2.8}
\end{eqnarray}
so we have
\begin{eqnarray}
&& \frac{d}{dt}P_n=e^{Q_{n-1}-Q_n}+c_1,\label{U1}\\
&& \frac{d}{dt}Q_n=e^{P_n-P_{n+1}+Q_{n-1}-Q_{n+1}}+c_2,\label{V1}
\end{eqnarray}
Here $c_1, c_2$ are two arbitrary constants. We now introduce the
new potentials $u_n=P_{n,t}-c_1, v_n=Q_{n,t}-c_2$ and differentiate
Eqs. \eqref{U1} and \eqref{V1} to find
\begin{align}
\frac{d}{dt}u_n & =e^{Q_{n-1}-Q_n}(v_{n-1}-v_n)=u_n(v_{n-1}-v_n),\label{u1}\\
\frac{d}{dt}v_n & =e^{P_n-P_{n+1}+Q_{n-1}-Q_{n+1}}(u_n-u_{n+1}
+v_{n-1}-v_{n+1})\nonumber\\
&=v_n(u_n-u_{n+1} +v_{n-1}-v_{n+1}),\label{v1}
\end{align}
which is the original nonlinear RT lattice \eqref{rt1} and
\eqref{rt2}. So \eqref{btU1}-\eqref{btU2} provide a Lax pair for RTL
\eqref{rt1}-\eqref{rt2}.

Using (\ref{bt21})-(\ref{bt24}), we can obtain a nontrivial solution
from the trivial solution. It is easy to check that $F'=G'=1$ is a
solution of \eqref{brt1}-\eqref{brt2}. Substituting this trivial
solution into the BT (\ref{bt21})-(\ref{bt24}), we have
\begin{eqnarray}
&& F_t-\lambda F_{n-1}+\mu F_n=0,\\
&& G_t-\lambda G_{n-1}+\mu G_n=0,\\
&& G_{n+1}-F_n-\lambda G_n=0,\\
&& \lambda G_{n-1}+G_n+\gamma F_n=0,
\end{eqnarray}
When we take $\lambda=2, \mu=3,\gamma=3$ we can obtain the following
solution of \eqref{brt1}-\eqref{brt2}
\begin{eqnarray}
&& F_n=(2/3)^{n+1}-2(2/3)^n-e^{-t},\quad G_n=(2/3)^n+e^{-t}.
\end{eqnarray}
In Fig. 1, we plot the one-soliton solution for $u$ and $v$
respectively.
\begin{figure}[htp]\label{fig:1}
\begin{center}
\begin{minipage}{0.4\textwidth}
\includegraphics[width=6cm]{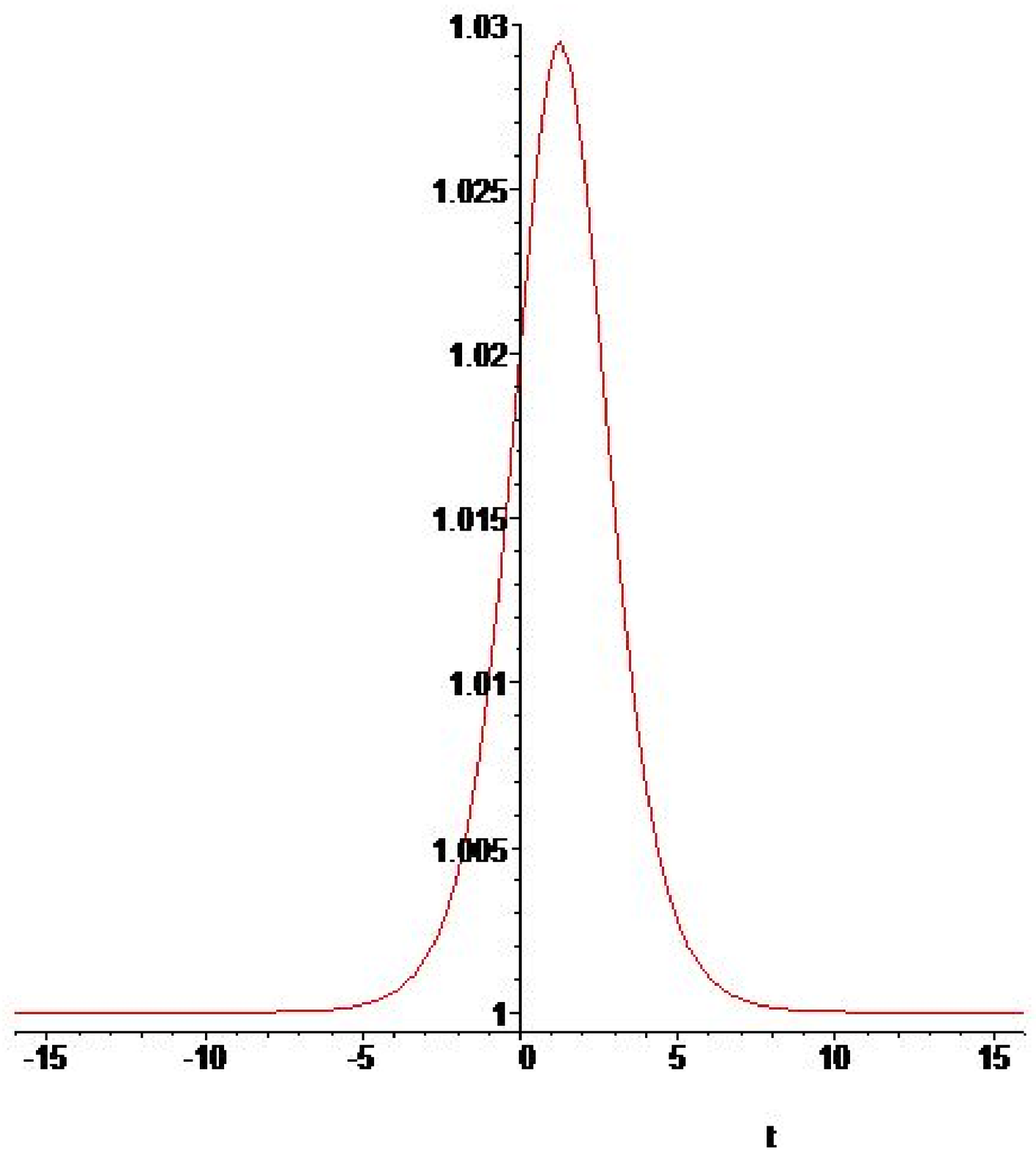}
\centerline{ $(a)\quad u_n$} 
\end{minipage}
\begin{minipage}{0.4\textwidth}
\includegraphics[width=6cm]{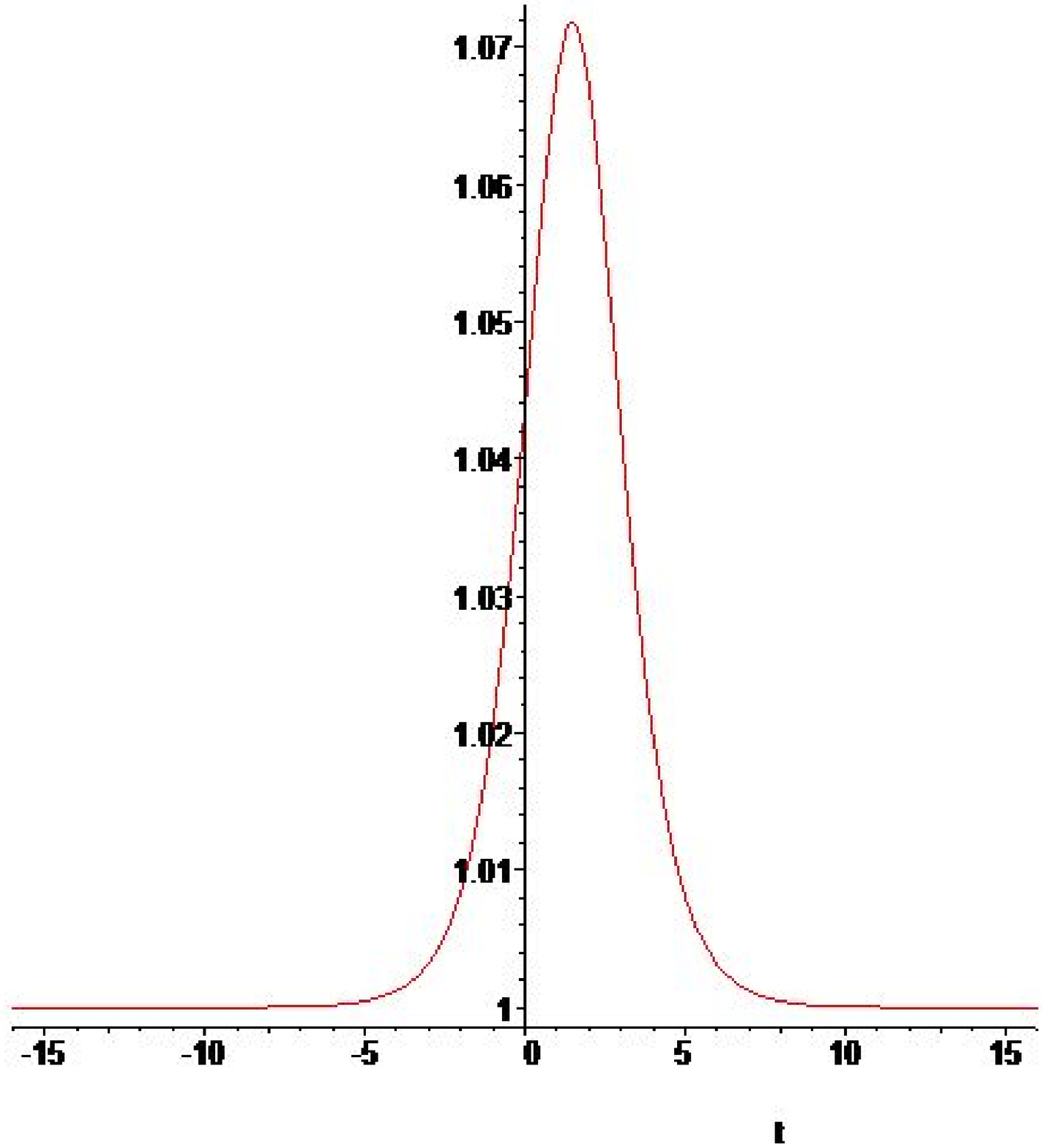}
\centerline{ $(b) \quad v_n$}
\end{minipage}\hspace{.05cm}
\caption{(Color online) The one-soliton solution for
\eqref{rt1}-\eqref{rt2}: $n=4.$}
\end{center}
\end{figure}

\section{Integrable full-discrete version of the RTL}
\setcounter{equation}{0} We propose the following bilinear equations
\begin{eqnarray}
&& [-\frac{1}{\epsilon}\sinh(\epsilon
D_t)e^{-\frac{D_m+D_n}{2}}-e^{-\frac{D_m+D_n}{2}+\epsilon
D_t}+e^{\frac{D_m-D_n}{2}-\epsilon D_t}]f\cdot f=0,\label{fk1}\\
&& [-\frac{1}{\epsilon}\sinh(\epsilon
D_t)e^{-\frac{D_m}{2}}+e^{-\frac{D_m}{2}-\epsilon
D_t}-e^{\frac{D_m}{2}-D_n-\epsilon D_t}]f\cdot f=0.\label{fk2}
\end{eqnarray}
In the continuum limit as $\epsilon \rightarrow 0$, the system
(\ref{fk1}) and (\ref{fk2}) reduces to the lattice (\ref{brt3}) and
(\ref{brt4}). The system (\ref{fk1}) and (\ref{fk2}) therefore
provides a fully discrete version of (\ref{brt3}) and (\ref{brt4}).

Note that
\begin{eqnarray}
F=f_{m+\frac{1}{2}},\quad G=f_{m-\frac{1}{2}},
\end{eqnarray}
and denote the variable $t$ as $k$, Eqs. \eqref{fk1}-\eqref{fk2} can
be rewritten as
\begin{align}
& (-\frac{1}{2\epsilon}-1)G_{n}^{k+1}F_{n+1}^{k-1}
+\frac{1}{2\epsilon}G_n^{k-1}F_{n+1}^{k+1}+G_{n+1}^{k+1}F_n^{k-1}=0,\label{f3}\\
&(\frac{1}{2\epsilon}+1)G_{n}^{k-1}F_n^{k+1}-\frac{1}{2\epsilon}G_n^{k+1}F_n^{k-1}
-G_{n+1}^{k+1}F_{n-1}^{k-1}=0.\label{f4}
\end{align}

With the dependent variable transformation
\begin{eqnarray}
u_{n}^{k}=\frac{G_{n-1}^{k}}{F_{n}^{k}},\quad
v_{n}^{k}=\frac{F_{n}^{k}}{G_{n}^{k}}, \label{vt}
\end{eqnarray}
the bilinear equations (\ref{f3})-(\ref{f4}) are put in the
following nonlinear form:
\begin{align}
&(-\frac{1}{2\epsilon}-1)u_{n+1}^{k+1}v_{n+1}^{k+1}+u_{n+1}^{k-1}v_{n}^{k-1}
+\frac{1}{2\epsilon}u_{n+1}^{k-1}v_{n+1}^{k+1}=0,\label{nf3}\\
& (\frac{1}{2\epsilon}+1)v_{n}^{k+1}u_{n+1}^{k+1}v_{n+1}^{k+1}
-\frac{1}{2\epsilon}v_{n}^{k-1}u_{n+1}^{k+1}v_{n+1}^{k+1}-
u_{n}^{k-1}v_{n-1}^{k-1}v_{n}^{k-1}=0.\label{nf4}
\end{align}
Set $t=k\epsilon$. By using the Taylor formula,
\begin{eqnarray}
u_n^{k+1}=u_n((k+1)\epsilon)=u_n(t+\epsilon)=u_n(t)+
u'_n(t)\epsilon+\mathcal{O}(\epsilon^2)
\end{eqnarray}
and after careful calculations, we finally obtain that as $\epsilon
\rightarrow 0$ the continuum limits of the fully discrete Eqs.
\eqref{nf3} and \eqref{nf4} are given by
\begin{align}
& u_n'=u_n\left(\frac{v_{n-1}}{v_n}-1\right),\\
& v_n'=v_n\left(\frac{u_nv_{n-1}}{u_{n+1}v_{n+1}}-1\right).
\end{align}
Setting $d_n=(\ln u_n)'+1$ and $c_n=(\ln v_n)'+1$, then we get
\begin{align}
& d_n'=d_n(c_{n-1}-c_n),\\
& c_n'=c_n(c_{n-1}-c_{n+1}+d_n-d_{n+1}),
\end{align}
which is just the RT system \eqref{rt1} and \eqref{rt2}. Therefore
the system (\ref{nf3}) and (\ref{nf4}) provides a fully discrete 
version of the potential RT system. 

When we take the dependent variable transformation
\begin{align}\label{fdv}
U_n^k=\frac{F_n^{k-1}G_{n-1}^{k+1}}{F_n^{k+1}G_{n-1}^{k-1}}=\frac{u_{n}^{k+1}}{u_{n}^{k-1}},\quad 
V_n^k=\frac{F_n^{k+1}G_{n}^{k-1}}{F_n^{k-1}G_{n}^{k+1}}=\frac{v_{n}^{k+1}}{v_{n}^{k-1}},
\end{align}
we can deduce from (\ref{fk1})-(\ref{fk2}) that
\begin{align}
& (1+\frac{1}{2\epsilon})(U_{n}^{k+1}-U_{n}^{k-1})=\frac{v_{n-1}^k}{v_{n}^{k+2}}-\frac{v_{n-1}^{k-2}}{v_{n}^{k}},\label{fd1}\\
& (1+\frac{1}{2\epsilon})(V_{n}^{k+1}-V_{n}^{k-1})=\frac{u_n^kv_{n-1}^k}{u_{n+1}^{k+2}v_{n+1}^{k+2}}
-\frac{u_n^{k-2}v_{n-1}^{k-2}}{u_{n+1}^{k}v_{n+1}^{k}}.\label{fd2}
\end{align}
One can check that in the continuum limit $\epsilon\rightarrow 0$, Eqs. \eqref{fd1}-\eqref{fd2} reduce to 
the RTL equation \begin{align}
& U_n'=U_n(V_{n-1}-V_n),\\
& V_n'=V_n(V_{n-1}-V_{n+1}+U_n-U_{n+1}).
\end{align}
Hence equations \eqref{fdv}-\eqref{fd2} give a fully discrete
version of the  RT system (\ref{rt1}) and (\ref{rt2}). 
\begin{remark}
We remark here that the full discrete analogue of RTL obtained here is different
from \eqref{drt}. Suris obtained dRTL by using Hamiltonian method,
while here we start from Hirota bilinear method. What interesting
for integrable discretization is that one integrable system may have
some different integrable discrete analogues. Among these integrable
discrete versions, some are linked each other by appropriate
transformations. Others may have no direct relation. In our case, we have not found the direct 
link between our full discrete RTL to \eqref{drt}. 
\end{remark}

For simplicity
we take $\epsilon=1$ in the following. In this case, the
bilinear system (\ref{fk1})-(\ref{fk2}) and nonlinear equations
\eqref{nf3}-\eqref{nf4} become
\begin{eqnarray}
&& (e^{\frac{D_m-D_n}{2}-D_k}-\frac{3}{2}e^{-\frac{D_m+D_n}{2}+D_k}
+\frac{1}{2}e^{-\frac{D_m+D_n}{2}-D_k})f\cdot
f=0,\label{full1}\\
&& (e^{\frac{D_m}{2}-D_n-D_k}+\frac{1}{2}e^{-\frac{D_m}{2}+D_k}
-\frac{3}{2}e^{-\frac{D_m}{2}-D_k})f\cdot f=0,\label{full2}
\end{eqnarray}
and
\begin{eqnarray}
&& u_{n+1}^{k-1}v_{n}^{k-1}-\frac{3}{2}u_{n+1}^{k+1}v_{n+1}^{k+1}
+\frac{1}{2}u_{n+1}^{k-1}v_{n+1}^{k+1}=0,\label{nf1}\\
&&
u_{n}^{k-1}v_{n-1}^{k-1}v_{n}^{k-1}+\frac{1}{2}v_{n}^{k-1}u_{n+1}^{k+1}v_{n+1}^{k+1}-
\frac{3}{2}v_{n}^{k+1}u_{n+1}^{k+1}v_{n+1}^{k+1}=0.\label{nf2}
\end{eqnarray}
respectively.

We will
show that (\ref{nf1})-(\ref{nf2}) is integrable by exhibiting its BT
and Lax pair. Concerning the bilinear equations
(\ref{full1})-(\ref{full2}), we have the following result:
\begin{prop}
The bilinear equations (\ref{full1})-(\ref{full2}) have the
B\"acklund transformation
\begin{eqnarray}
&& (e^{\frac{D_m}{2}}+\gamma e^{-\frac{D_m}{2}}-2\mu
e^{\frac{D_m}{2}-D_n})f\cdot g=0,\label{bf1}\\
&& (e^{\frac{D_n}{2}+D_k}+\lambda e^{-\frac{D_n}{2}-D_k}+\mu
e^{-\frac{D_n}{2}+D_k})f\cdot g=0,\label{bf2}\\
&& (e^{\frac{D_m+D_n}{2}}+\frac{2}{3}\mu
e^{\frac{D_m-D_n}{2}}+\alpha e^{-\frac{D_m+D_n}{2}})f\cdot
g=0.\label{bf3}
\end{eqnarray}
where $\lambda, \mu, \gamma$ and $\alpha$ are arbitrary constants.
\end{prop}
{\bf Proof.} Let $f$ be a solution of equations
\eqref{full1}-\eqref{full2}. If it can be shown that $g$ given by
\eqref{bf1}-\eqref{bf3} satisfies
\begin{eqnarray*}
&&  P_1\equiv
(e^{\frac{D_m-D_n}{2}-D_k}-\frac{3}{2}e^{-\frac{D_m+D_n}{2}+D_k}
+\frac{1}{2}e^{-\frac{D_m+D_n}{2}-D_k})g\cdot
g=0,\\
&&  P_2\equiv
(e^{\frac{D_m}{2}-D_n-D_k}+\frac{1}{2}e^{-\frac{D_m}{2}+D_k}
-\frac{3}{2}e^{-\frac{D_m}{2}-D_k})g\cdot g=0,
\end{eqnarray*}
then the equations \eqref{bf1}-\eqref{bf3} provide a BT for
\eqref{full1}-\eqref{full2}.

In fact, by using \eqref{bf1}-\eqref{bf3} and the bilinear
identities, one can see that

\begin{align*}
&-[e^{-\frac{D_m+D_n}{2}-D_k}f\cdot f]P_1 \\
&\quad\equiv
[(e^{\frac{D_m-D_n}{2}-D_k}-\frac{3}{2}e^{-\frac{D_m+D_n}{2}+D_k}
+\frac{1}{2}e^{-\frac{D_m+D_n}{2}-D_k})f
\cdot f][e^{-\frac{D_m+D_n}{2}-D_k}g\cdot g]\\
&\qquad \qquad \qquad
-[e^{\frac{D_m-D_n}{2}-D_k}-\frac{3}{2}e^{-\frac{D_m+D_n}{2}+D_k}
+\frac{1}{2}e^{-\frac{D_m+D_n}{2}-D_k})g \cdot g]
[e^{-\frac{D_m+D_n}{2}-D_k}f\cdot f]\\
&\quad =2\sinh(\frac{D_m}{2})(e^{-\frac{D_n}{2}-D_k}f\cdot g)\cdot
(e^{\frac{D_n}{2}+D_k}f\cdot
g)-3\sinh(D_k)(e^{-\frac{D_n+D_m}{2}}f\cdot g)\cdot
(e^{\frac{D_m+D_n}{2}}f\cdot g)\\
& \quad =-2\mu \sinh(\frac{D_m}{2})(e^{-\frac{D_n}{2}-D_k} f\cdot
g)\cdot (e^{-\frac{D_n}{2}+D_k}f\cdot
g)-3\sinh(D_k)(e^{-\frac{D_n+D_m}{2}}f\cdot g)\cdot
(e^{\frac{D_m+D_n}{2}}f\cdot g)\\
&\quad=2\mu\sinh(D_k)[e^{\frac{D_m-D_n}{2}}f\cdot g] \cdot
(e^{-\frac{D_m+D_n}{2}}f\cdot g)
-3\sinh(D_k)(e^{-\frac{D_n+D_m}{2}}f\cdot g)\cdot
(e^{\frac{D_m+D_n}{2}}f\cdot g)\\
&\quad =-3\sinh(D_k)(e^{-\frac{D_m+D_n}{2}}f\cdot g)\cdot [(
e^{\frac{D_m+D_n}{2}}+\frac{2}{3}\mu e^{\frac{D_m-D_n}{2}})f\cdot g]\\
&\quad =0\\
\end{align*}
and similarly that
\begin{align*}
&-[e^{-\frac{D_m}{2}-D_k} f\cdot f]P_2 \\
&\quad\equiv
[(e^{\frac{D_m}{2}-D_n-D_k}+\frac{1}{2}e^{-\frac{D_m}{2}+D_k}
-\frac{3}{2}e^{-\frac{D_m}{2}-D_k})f\cdot f][e^{-\frac{D_m}{2}-D_k}g\cdot g]\\
&\qquad\qquad\qquad\qquad -
[(e^{\frac{D_m}{2}-D_n-D_k}+\frac{1}{2}e^{-\frac{D_m}{2}+D_k}
-\frac{3}{2}e^{-\frac{D_m}{2}-D_k})g\cdot g][e^{-\frac{D_m}{2}-D_k}f\cdot f]\\
&\quad =2\sinh(\frac{D_m-D_n}{2})(e^{-\frac{D_n}{2}-D_k}f\cdot
g)\cdot (e^{\frac{D_n}{2}+D_k}f\cdot
g)+\sinh(D_k)(e^{-\frac{D_m}{2}}f\cdot g)\cdot
(e^{\frac{D_m}{2}}f\cdot g)\\
&\quad =-2\mu\sinh(\frac{D_m-D_n}{2})(e^{-\frac{D_n}{2}-D_k}f\cdot
g)\cdot (e^{-\frac{D_n}{2}+D_k}f\cdot g)+\sinh(D_k)
(e^{-\frac{D_m}{2}}f\cdot g)\cdot (e^{\frac{D_m}{2}}f\cdot g)\\
&\quad =2\mu \sinh(D_k)(e^{\frac{D_m}{2}-D_n}f\cdot g)\cdot
(e^{-\frac{D_m}{2}}f\cdot g)+\sinh(D_k)(e^{-\frac{D_m}{2}}f\cdot
g)\cdot(e^{\frac{D_m}{2}}f\cdot g)\\
&\quad =\sinh(D_k)(e^{-\frac{D_m}{2}}f\cdot g)\cdot
[(e^{\frac{D_m}{2}}-2\mu e^{\frac{D_m}{2}-D_n})f\cdot g]\\
&\quad=0.
\end{align*}
This completes the proof of proposition $4.1$. Again we can obtain
from Proposition 3.1 that the bilinear fully discrete RT lattice
\eqref{f3}-\eqref{f4} ($\epsilon=1$) has the following BT:
\begin{eqnarray}
&& F_n G_n'+\gamma G_n F_n'-2\mu F_{n-1}G'_{n+1}=0,\label{fbt1}\\
&& F_{n+1}G'_{n}+\frac{2}{3}\mu F_{n}G'_{n+1}+\alpha
G_{n}F'_{n+1}=0,\\
&& F_{n+1,k+1}F_{n,k-1}'+\lambda F_{n,k-1}F_{n+1,k+1}'+\mu
F_{n,k+1}F_{n+1,k-1}'=0,\\
&& G_{n+1,k+1}G_{n,k-1}'+\lambda G_{n,k-1}G_{n+1,k+1}'+\mu
G_{n,k+1}G_{n+1,k-1}'=0.\label{fbt4}
\end{eqnarray}
Using (\ref{fbt1})-(\ref{fbt4}), we can obtain a nontrivial solution
from the trivial solution. It is easy to check that $F'=G'=1$ is a
solution of \eqref{f3}-\eqref{f4}. Substituting this trivial
solution into the BT (\ref{fbt1})-(\ref{fbt4}), we have
\begin{eqnarray}
&& F_n+\gamma G_n-2\mu F_{n-1}=0,\label{slu1}\\
&& F_{n+1}+\frac{2}{3}\mu F_n+\alpha G_n=0,\\
&& F_{n+1}^{k+1}+\lambda F_n^{k-1}+\mu F_n^{k+1}=0,\\
&& G_{n+1}^{k+1}+\lambda G_n^{k-1}+\mu G_n^{k+1}=0.\label{slu2}
\end{eqnarray}
When we take $\alpha=11\gamma, \mu=1, \lambda=-100$, we get the
following solution of \eqref{slu1}-\eqref{slu2}
\begin{eqnarray}
&& F_n=c_15^k3^n+c_2(\sqrt{12})^k(\frac{22}{3})^n,\\
&&G_n=2\left[c_15^k3^{n-1}+c_2(\sqrt{12})^k(\frac{22}{3})^{n-1}\right]
-c_15^k3^n-c_2(\sqrt{12})^k(\frac{22}{3})^n.
\end{eqnarray}
$c_1$ and $c_2$ are arbitrary constants. In Fig. 2, we plot the
one-soliton solution for $U_n^k$ and $V_n^k$ by use of \eqref{fdv}
with parameters selected as $c_1=2,c_2=1$ and $k=5$.
\begin{figure}[htp]\label{fig:2}
\begin{center}
\begin{minipage}{0.4\textwidth}
\includegraphics[width=6cm]{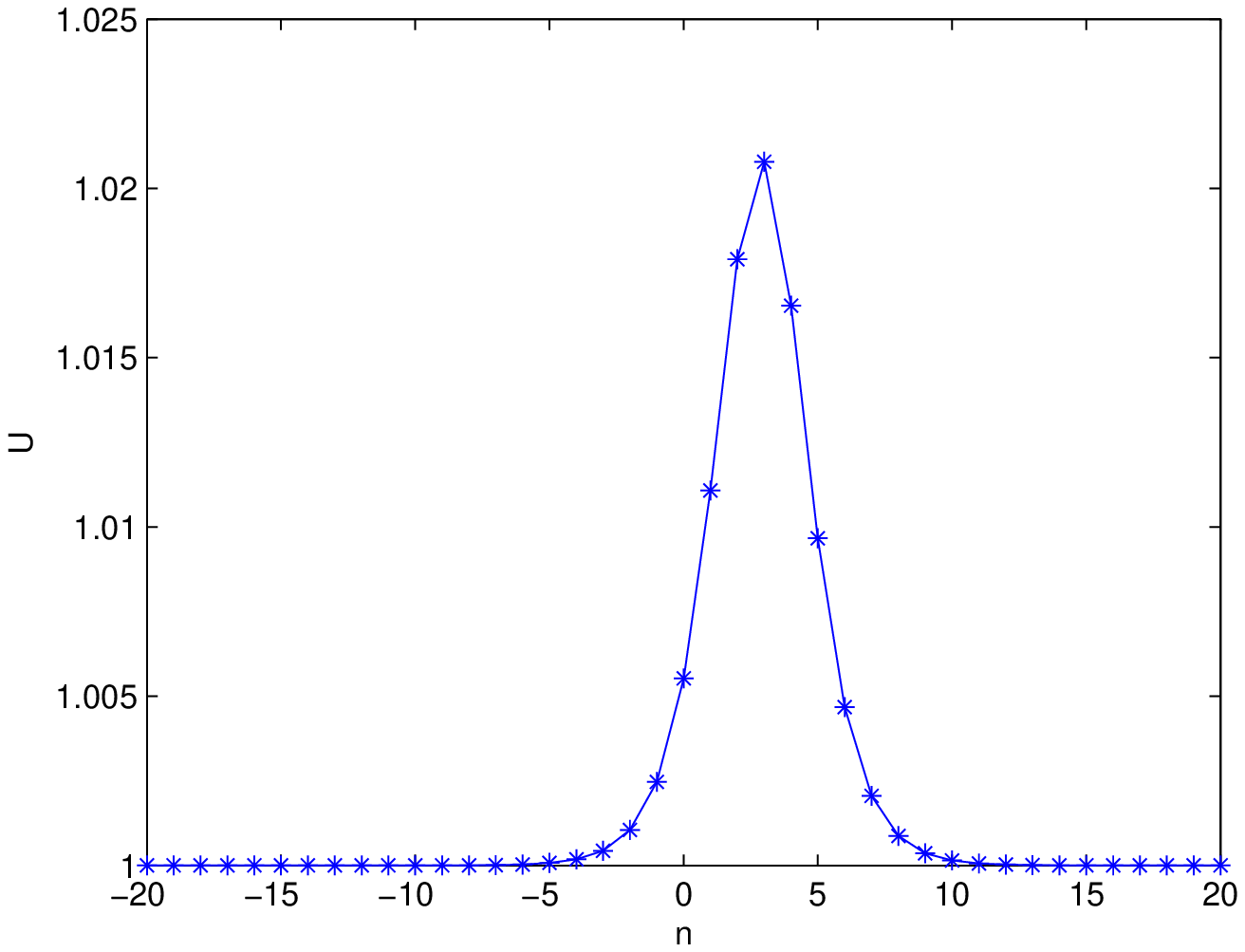}
\centerline{ $(a) \quad U_n^k$} 
\end{minipage}
\begin{minipage}{0.4\textwidth}
\includegraphics[width=6cm]{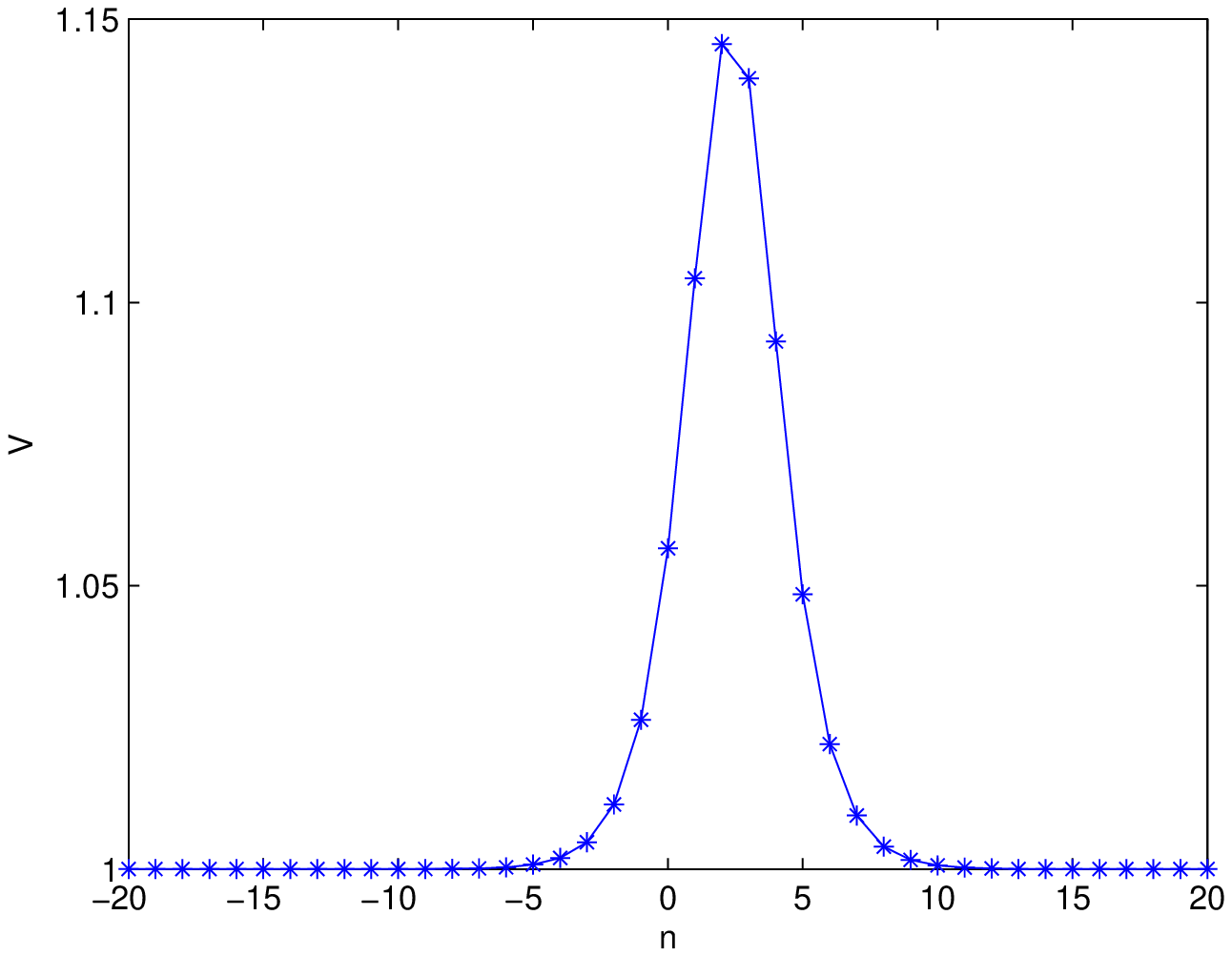}
\centerline{ $(b)\quad  V_n^k $}
\end{minipage}\hspace{.05cm}
\caption{(Color online) The one-soliton solution for full-discrete
system (\ref{fdv})-(\ref{fd2}).}
\end{center}
\end{figure}

In the following, we derive a Lax pair for the fully discrete RTL
(\ref{nf1})-(\ref{nf2}). Let
\begin{eqnarray}
&& \frac{F'_n}{G'_n}=v_n,\quad \frac{G'_{n-1}}{F'_n}=u_n,\quad
\frac{F}{F'}=\Phi,\quad \frac{G}{G'}=\Psi.
\end{eqnarray}
From  Eqs. (\ref{fbt1})-(\ref{fbt4}) we get
\begin{eqnarray}
&& \Phi_nu_{n+1}v_{n+1}+\gamma\Psi_nu_{n+1}v_{n+1} -2\mu\Phi_{n-1}u_nv_{n-1} =0,\label{lax1}\\
&& \Phi_{n+1}v_{n+1}+\alpha \Psi_{n}v_{n+1}
+\frac{2}{3}\mu \Phi_{n}v_{n}=0,\label{lax2}\\
&&\Phi_{n+1}^{k+1}u_{n+1}^{k-1}v_{n}^{k-1}+\lambda
\Phi_{n}^{k-1}u_{n+1}^{k-1}v_{n}^{k-1}+\mu
\Phi_{n}^{k+1}u_{n+1}^{k+1}v_{n}^{k+1}=0,\label{lax3}\\
&&\Psi_{n+1}^{k+1}u_{n+1}^{k-1}v_{n+1}^{k-1}+\lambda
\Psi_{n}^{k-1}u_{n+1}^{k-1}v_{n+1}^{k-1}+\mu
\Psi_{n}^{k+1}u_{n+1}^{k+1}v_{n+1}^{k+1}=0.\label{lax4}
\end{eqnarray}
In order to obtain a Lax pair for \eqref{nf1} and \eqref{nf2}, we
first use \eqref{lax2} to express $\Psi$ in \eqref{lax4} in terms of
$\Phi$ and then to write  $\Phi_{n+2}$ and $\Phi_{n+1}$ as
expressions of $\Phi_{n}$. After some calculations we find
\begin{eqnarray}
&& -2 u_{n+1}^{k-1} u_{n+2}^{k-1} v_{n+1}^{k-1} v_{n+1}^{k+1}+2
u_{n+1}^{k-1}
u_{n+2}^{k-1} v_{n}^{k-1}v_{n+2}^{k+1}\nonumber \\
&&\qquad -3 u_{n+1}^{k+1} u_{n+2}^{k-1} v_{n+1}^{k+1}
v_{n+2}^{k+1}+3 u_{n+1}^{k-1} u_{n+2}^{k+1} v_{n+1}^{k+1}
v_{n+2}^{k+1}=0,
\end{eqnarray}
or
\begin{eqnarray}
&&\frac{2 u_{n+2}^{k-1} v_{n+1}^{k-1}-3 u_{n+2}^{k+1}
v_{n+2}^{k+1}}{u_{n+2}^{k-1} v_{n+2}^{k+1}}
-\frac{2 u_{n+1}^{k-1} v_{n}^{k-1}
-3 u_{n+1}^{k+1} v_{n+1}^{k+1}}{u_{n+1}^{k-1} v_{n+1}^{k+1}}\nonumber      \\
&&=(E_{n}-1)\frac{2 u_{n+1}^{k-1} v_{n}^{k-1}-3 u_{n+1}^{k+1}
v_{n+1}^{k+1}}{u_{n+1}^{k-1} v_{n+1}^{k+1}}=0.
\end{eqnarray}
It follows that
\begin{eqnarray}
\frac{2 u_{n+1}^{k-1} v_{n}^{k-1}-3 u_{n+1}^{k+1}
v_{n+1}^{k+1}}{u_{n+1}^{k-1} v_{n+1}^{k+1}}=-c,
\end{eqnarray}
or equivalently,
\begin{eqnarray}
2 u_{n+1}^{k-1} v_{n}^{k-1}-3 u_{n+1}^{k+1} v_{n+1}^{k+1}+c
u_{n+1}^{k-1} v_{n+1}^{k+1}=0.
\end{eqnarray}
Setting $u_{n}^{k}\rightarrow c^{-k/2}u_{n}^{k},
v_{n}^{k}\rightarrow c^{k/2}v_{n}^{k}$, we have
\begin{eqnarray}
2 u_{n+1}^{k-1} v_{n}^{k-1}-3 u_{n+1}^{k+1} v_{n+1}^{k+1}+
u_{n+1}^{k-1} v_{n+1}^{k+1}=0,
\end{eqnarray}
which coincides with \eqref{nf1}. Similarly, combining \eqref{lax1}
and \eqref{lax3}, we get
\begin{eqnarray}
&&-2u_{n+1}^{k-1}u_{n+1}^{1+k}v_{n}^{k-1}v_{n}^{1+k}v_{n+1}^{k-1}
+2u_{n}^{k-1}u_{2+n}^{1+k}v_{n-1}^{k-1}v_{n}^{k-1}v_{2+n}^{1+k}
\nonumber  \\
&&\qquad -u_{1+n}^{1+k} u_{2+n}^{1+k} v_{n}^{1+k} v_{n+1}^{k-1}
v_{2+n}^{1+k}+u_{1+n}^{1+k} u_{2+n}^{1+k} v_{n}^{k-1} v_{1+n}^{1+k}
v_{2+n}^{1+k}=0,
\end{eqnarray}
or
\begin{eqnarray}
&&\frac{\left(2 u_{n+1}^{k-1} v_{n}^{k-1}+u_{2+n}^{1+k}
v_{2+n}^{1+k}\right) v_{n+1}^{k-1}}{u_{2+n}^{1+k}
v_{2+n}^{1+k}v_{n+1}^{k+1}}-\frac{\left(2 u_{n}^{k-1}
v_{n-1}^{k-1}+u_{1+n}^{1+k} v_{1+n}^{1+k}\right)
v_{n}^{k-1}}{u_{1+n}^{1+k}v_{n+1}^{k+1} v_{n}^{1+k}}\nonumber   \\
&&=(E_n-1)\frac{\left(2 u_{n}^{k-1} v_{n-1}^{k-1}+u_{1+n}^{1+k}
v_{1+n}^{1+k}\right)v_{n}^{k-1}}{u_{1+n}^{1+k}v_{n+1}^{k+1}
v_{n}^{1+k}}=0.
\end{eqnarray}
We then have
\begin{eqnarray}
\frac{\left(2 u_{n}^{k-1} v_{n-1}^{k-1}+u_{1+n}^{1+k}
v_{1+n}^{1+k}\right)v_{n}^{k-1}}{u_{1+n}^{1+k}v_{n+1}^{k+1}
v_{n}^{1+k}}=3c,
\end{eqnarray}
or
\begin{eqnarray}
\left(2 u_{n}^{k-1} v_{n-1}^{k-1}+u_{n+1}^{k+1}
v_{n+1}^{k+1}\right)v_{n}^{k-1}-3c u_{n+1}^{k+1}v_{n+1}^{k+1}
v_{n}^{k+1}=0.
\end{eqnarray}
Setting $u_{n}^{k}\rightarrow c^{k/2}u_{n}^{k}, v_{n}^{k}\rightarrow
c^{-k/2}v_{n}^{k}$, we arrive at
\begin{eqnarray}
\left(2 u_{n}^{k-1} v_{n-1}^{k-1}+u_{n+1}^{k+1}
v_{n+1}^{k+1}\right)v_{n}^{k-1}-3 u_{n+1}^{k+1}v_{n+1}^{k+1}
v_{n}^{k+1}=0.
\end{eqnarray}
which checks with \eqref{nf2}. So Eqs.\eqref{lax1}-\eqref{lax4}
constitute the Lax pair for (\ref{nf1})-(\ref{nf2}).

\section{Conclusion}
In this paper we present a bilinear form for a integrable lattice
related to the well-known relativistic Toda lattice. Through
Hirota's bilinear integrable discretization method a fully discrete
version of the lattice is obtained. Bilinear BT and Lax pair for
both the original and the fully discrete lattice are investigated.

\section*{\bf Acknowledgements}
The authors would like to express their thanks to the referees for valuable advice. G.Yu and Y.Zhang are grateful to Xing-Biao Hu for helpful discussions. The work of L.V. is supported by a grant from the
National Science and Engineering Research Council (NSERC) of Canada.
G.Yu acknowledges a postdoctoral fellowship from the CRM of
Mathematical Physics Laboratory. G.Yu and Y.Zhang are also supported by the
National Natural Science Foundation of China (Grant no.11371251,11271362) and Chenguang Program (No.09CG08) sponsored by Shanghai Municipal
Education Commission and Shanghai Educational Development
Foundation.  \vskip .5cm

\end{document}